# Monitoring oligomerization dynamics of individual human neurotensin receptors 1 in living cells and in SMALP nanodiscs

Lukas Spantzel[a], Iván Pérez[a], Thomas Heitkamp[a], Anika Westphal[b], Stefanie Reuter[b], Ralf Mrowka[b], Michael Börsch*[a,c]

[a]Single-Molecule Microscopy Group, Jena University Hospital, Nonnenplan 2 - 4, 07743 Jena;
[b]Experimental Nephrology Group, Jena University Hospital, Nonnenplan 2 - 4, 07743 Jena;
[c]Abbe Center of Photonics (ACP) Jena, Germany

## ABSTRACT

The human neurotensin receptor 1 (NTSR1) is a G protein-coupled receptor. The receptor is activated by a small peptide ligand neurotensin. NTSR1 can be expressed in HEK cells by stable transfection. Previously we used the fluorescent protein markers mRuby3 or mNeonGreen fused to NTSR1 for EMCCD-based structured illumination microscopy (SIM) in living HEK cells. Ligand binding induced conformational changes in NTSR1 which triggered the intracellular signaling processes. Recent single-molecule studies revealed a dynamic monomer/dimer equilibrium of this receptor in artificial lipid bilayers. Here we report on the oligomerization state of human NTSR1 from living cells by trapping them into lipid nanodiscs. Briefly, SMALPs (styrene-maleic acid copolymer lipid nanoparticles) were produced directly from the plasma membranes of living HEK293T FlpIn cells. SMALPs with a diameter of 15 nm were soluble and stable. NTSR1 in SMALPs were analyzed by single-molecule intensity measurements one membrane patch at a time using a custom-built confocal anti-Brownian electrokinetic trap (ABEL trap) microscope. We found oligomerization changes before and after stimulation of the receptor with its ligand neurotensin.

**Keywords:** Neurotensin receptor 1, SMALP, single-molecule detection, confocal, ABEL trap, FLIM

## 1. INTRODUCTION

The human neurotensin receptor 1 (NTSR1) is a G protein-coupled receptor (GPCR)[1]. GPCRs are membrane proteins that can recognize extracellular ligands, respond to mechanical stimuli, light, or pheromones, and convert them into intracellular signals. Understanding the molecular mechanisms of these receptors builds the basis for therapeutic drug development, as one-third of all drugs used today are targeting GPCRs[2, 3]. Only 30 of the 750 identified GPCRs are currently targets for clinical drugs[4, 5]. This increases the possibility of new drug discoveries in this field. A central topic about the first steps in the signaling pathway is the role of monomeric *versus* oligomeric GPCRs as the active signal transmitters.

NTSR1 belongs to the β group of class A GPCRs. It participates in modulating dopaminergic systems, analgesia, and inhibition of food intake. It is localized in the brain and the intestine[1, 6]. The structure of this receptor has been solved first by the group of R. Grisshammer in 2012[1]. In the last years, new NTSR1 structures became available improving the understanding of the method of operation[7-13]. NTSR1 is stimulated by the ligand neurotensin (NTS). NTS is a 13-amino acid peptide and is modulating a variety of functions in the central and peripheral nervous system[14-20].

Recent single-molecule studies revealed a dynamic monomer/dimer equilibrium of the purified receptor in artificial lipid bilayers[3, 6, 21-23]. For comparison with the native lipid environment, we started to investigate the oligomerization state of NTSR1 in living human cells. Therefore we constructed a variety of receptor mutants with either mNeonGreen[24] or mRuby3[25] as fluorescent marker proteins fused to the intracellular C-terminus of human NTSR1. The receptor mutants were stably expressed in HEK293T FlpIn cells lacking a native NTSR1[6, 26]. Widefield imaging using structured illumination revealed a homogeneous localization of the receptors at the plasma membrane.

..................................................................................................................................................
*michael.boersch@med.uni-jena.de; https://www.uniklinikum-jena.de/singlemoleculemicroscopy/en/

Addition of the ligand NTS to the HEK293T cells resulted in the formation of receptor aggregates as bright fluorescent clusters within minutes, confirming a functional ligand binding site of the receptor mutants and indicating an activation of the NTSR1 for possible signaling and internalization. Using fluorescent NTS derivatives, plasma membrane localization and subsequent clustering of NTSR1 had been visualized by confocal microscopy previously[27].

However, the homogeneous distribution of our NTSR1 mutants in the membrane of transfected cells does not allow to discriminate monomers from dimers of the receptor using intensity-based widefield fluorescence microscopies. Therefore, we decided to extract NTSR1 with its surrounding native lipid environment from the membrane. All cellular membranes were fragmented into lipid nanoparticles by adding styrene-maleic acid copolymers (SMA, see review[28]). This detergent-free solubilization of membrane proteins avoids disruption and loss of the native lipid environment. The styrene-maleic acid lipid nanoparticle (SMALP) formation approach is based on the amphiphilic properties of the copolymers[29, 30], i.e., the styrene domains interact with the hydrophobic core of lipid bilayers and the maleic acid domains interact with the aqueous buffer simultaneously. As the result, water-soluble disc-shaped lipid nanostructures containing membrane proteins including NTSR1 are obtained. Depending on the polymer used, SMALP disc sizes can be adjusted between less than 10 nm and up to 25 nm. Being well-defined small soluble nanoparticles, fluorescence of individual SMALPs can by analyzed one after another by confocal single-molecule detection methods in solution.

One biochemical advantage of SMALP nanodiscs compared to artificial membrane vesicles (liposomes) is the unrestricted accessibility of both the extracellular as well as the intracellular side of the membrane proteins. A second advantage is the high negative charge of the nanoparticles due to the deprotonated maleic acid domains of the copolymer. High charges allow long trapping of individual SMALPs by a confocal Anti-Brownian Electrokinetic trap (ABEL trap). The ABEL trap for quantitative single nanoparticle spectroscopy in solution was invented by A. E. Cohen and W. E. Moerner[31-34]. Here, we analyzed the brightness of the NTSR1-SMALPs in our confocal ABEL trap[35-38]. We found different fluorescence intensity levels in SMALPs derived from HEK293T cells expressing NTSR1-mNeonGreen either before or after stimulation with its ligand neurotensin. According to the intensity distributions of these SMALPs, NTSR1 likely exists as a monomer but also as a dimer or as an oligomer in the native membranes of mammalian cells following the addition of neurotensin.

## 2. EXPERIMENTAL PROCEDURES

### 2.1 Stable expression of NTSR1-mNeonGreen in living HEK293T FlpIn cells

The G protein-coupled receptor NTSR1 consists of seven transmembrane helices, an extracellular N-terminal domain and an intracellular C-terminal domain. The fluorescent protein mNeonGreen was inserted at the C-terminus of the receptor yielding NTSR1-mNeonGreen (NTSR1-mNG in the following). Cloning, generation of a stable HEK293T FlpIn cell line and cell culture was achieved using the methods and protocols as described[26]. HEK293T FlpIn cells were split every 7 days when 90-100 % confluence was reached. Cell culture media was supplemented with 3 µg/ml puromycin (InvivoGen, San Diego, USA). As a control, our previously published HEK293T cell line expressing the mitochondrial $F_oF_1$-ATP synthase with mNeonGreen fused to the γ-subunit[39] was grown similarly except for using 100 µg/ml hygromycin B (Carl Roth, Karlsruhe, Germany).

### 2.2 Polarization-resolved confocal fluorescence lifetime imaging microscopy (FLIM)

HEK293T FlpIn cells expressing NTSR1-mNG were seeded into a microfluidic sample chamber (µ-slide VI$^{0.5}$, IBIDI, Germany). The chamber was filled with high-glucose Dulbecco's modified Eagle's medium (DMEM, Gibco) with 10 % fetal bovine serum (Biochrom, Germany) and appropriate antibiotics. Cells were allowed to grow for four days at 37 °C. The media was exchanged after 3 days. One hour before imaging, the growth media was removed and replaced by 100 µl PBS+ buffer (DPBS, ThermoFischer Scientific; supplemented with 0.9 mM $CaCl_2$ and 0.5 mM $MgCl_2$) to decrease the background fluorescence outside of the cells and to stop the expression of new NTSR1-mNG. Thus, the fluorescence contrast between plasma membrane and cytoplasm was improved. Confocal imaging was performed at 37 °C. For stimulation of NTSR1, a pre-heated amount of 10 µl comprising 5 µM neurotensin in PBS+ buffer was carefully added into the microfluidic sample chamber. Afterwards aggregation of NTSR1 in the plasma membrane happened within minutes.

Fluorescence lifetime imaging (FLIM) was recorded using a confocal STED microscope (Expertline, Abberior Instruments, Göttingen, Germany). Briefly, mNeonGreen was excited with 488 nm by a ps-pulsed laser with 40 MHz repetition rate. The excitation polarization mode was selected for either circularly polarized excitation or for linearly polarized excitation at different user-defined angles with respect to the x,y-plane of the microscope stage (IX83, Olympus, Japan). Cells were imaged using the 60x water immersion objective UPLSAPO60XW (NA 1.2, Olympus, Japan). Fluorescence emission was recorded simultaneously by two single photon counting avalanche photodiodes (APDs) selected for high time resolution (SPCM-AQRH-14-TR, Excelitas, Canada). A polarizing beam splitter cube separated the emission for time-resolved anisotropy imaging in the spectral range from 500 nm to 550 nm (AHF, Tübingen, Germany). In parallel to the Abberior FPGA counter electronics, time-correlated single photon counting (TCSPC) was achieved using a synchronized set of four SPC-150N electronic cards plus four routers (Becker&Hickl, Berlin, Germany). We used 21 ps time binning for a high time resolution of the fluorescence decays. FLIM images were analyzed using the Imspector 16.3 software (Abberior Instruments) for direct export into the SPCImage 8.6 NG software (Becker&Hickl).

**2.3 Preparation of SMALPs with NTSR1-mNG from HEK293T cells**

SMALPs were produced directly from living HEK293T FlpIn cells stably expressing the NTSR1-mNG mutant (Fig. 1). We have described our approach in detail previously[6]. Briefly, large flasks (75 cm$^2$) with adherent HEK293T FlpIn cells were used at a confluence of 70-90 %. Cells were washed with pre-warmed PBS+ buffer and either kept with PBS+ for 2 h at 37 °C, or, to induce NTSR1 clustering, 1 µM neurotensin was added after 1 h and kept for another hour, respectively. Afterwards cells were scraped of and separated from the buffer by centrifugation at 200 x g. The cell pellet was quickly mixed with a 2%-solution of the styrene-maleic acid copolymer (XIRAN SL25010 P20; Polyscope, Geleen, Netherlands) in 1 ml SMA buffer (20 mM HEPES, 1 mM EGTA, 1 mM Mg-acetate, pH 7.4 at room temperature) plus 5 U/ml freshly added benzonase. The solution was shaken for 1 h at 37 °C to induce SMALP formation. SMALPs were found in the supernatant after ultracentrifugation at 100,000 x g (1 h at 4 °C). SMALPs were stored at 4 °C in the dark and used for single-molecule spectroscopy within a few days. If necessary, SMALPS were diluted with SMA buffer. The principle of SMALP preparation from HEK cells is depicted in Fig. 1.

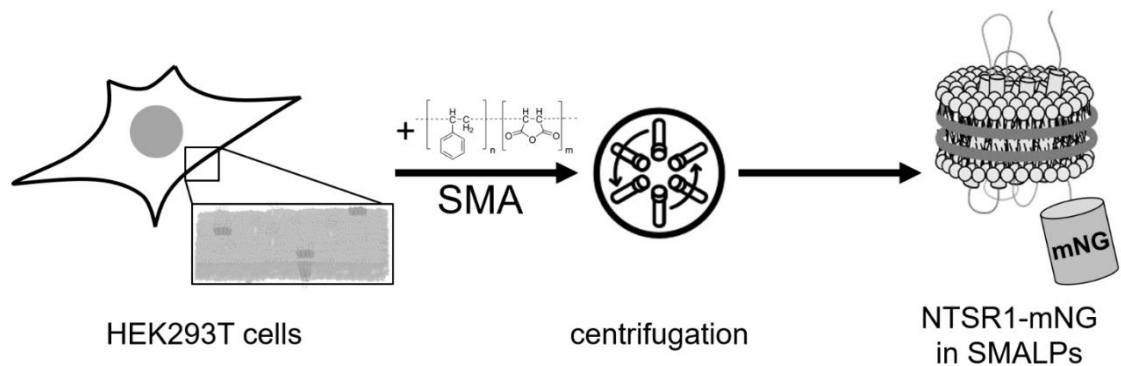

**Figure 1**: Schematic overview of the NTSR1-mNG in lipid nanodisc preparation from living HEK293T FlpIn cells to SMALPs using the styrene- maleic acid (SMA) copolymer approach. The model of NTSR1-mNG in a SMALP nanodisc is not to scale (modified from[30]).

**2.4 Production of SMALPs with F$_o$F$_1$-a-mNG-17-mNG-ATP synthase from *Escherichia coli***

As a reference for a permanent mNG dimer we constructed a fusion of two covalently linked mNeonGreen fluorophores to the membrane-embedded enzyme F$_o$F$_1$-ATP synthase from *Escherichia coli* (*E. coli*). The gene *atpB* (encoding the subunit *a* of the F$_o$F$_1$-ATP synthase from *E. coli*) was fused to a sequence encoding two mNeonGreen proteins connected by a 17 amino acid linker (mNG-17-mNG). To achieve this two *BamHI* restriction sites in pACWU-BHI[40] were removed by silent mutation. The remaining *BamHI* restriction site located between the C-terminus of *atpB* and the N-terminus of the gene for mNeonGreen site was converted to a *NdeI* restriction site. Furthermore, a *BamHI* restriction site was introduced to the N-terminus of the second mNeonGreen. All of the above was done by a modified QuikChange process[41]. With this, the plasmid pACWUBH1-NdeI-BamHI containing the sequence for *atpB-NdeI-mNG-BamHI* was created. The sequence *NdeI*-mNG-GTGASGGGGSGGSATAS-mNG-*BamHI* originating from pET15b-mNG-17-mNG was inserted with common cloning techniques. The final plasmid pACWU-TH2 was verified by DNA sequencing of the whole plasmid.

The $F_oF_1$-ATP synthase tagged with mNG-17-mNG was expressed in *E. coli* strain RA1 as described[42]. In contrast to our regular 10-L fermenter approach[43] the cell culture volume was scaled down to 1.6 liters. Sufficient oxygen supply during bacterial growth was ensured by vigorous shaking.

SMALPs with $F_oF_1$-a-mNG-17-mNG-ATP synthase were produced by dissolving the separated cell membrane of the bacteria. The membrane preparation was started by adding 20 ml lysis buffer (50 mM MOPS, 175 mM KCl, 10 mM $MgCl_2$, 0.2 mM EGTA, 0.2 mM DTT, pH 7.5 at 4 °C) to the cell pellet and stirring until homogenization. Afterwards another 20 ml lysis buffer was added. Shortly before cell disruption, a spatula tip of DNaseI and RNase (both AppliChem, Darmstadt, Germany), 0.1 mM PMSF and half of a tablet c0mplete protease inhibitor (Roche, Basel, Switzerland) were added. Cells were lysed by two passages through a PandaPlus 2000 cell homogenizer (GEA Niro Soavi, Italy) at 1000 bar. The cell lysate was centrifuged for 20 min at 25,000 x g, 4 °C, to remove cell debris and large particles. The supernatant was centrifuged for 2 h at 300,000 x g, 4 °C. The pellet containing the cell membrane was resuspended in 5 ml SMA buffer with freshly added 2 % XIRAN SL25010 P20 (Polyscope, Geleen, Netherlands). SMALP formation took place during vigorous shaking at 37 °C for 1 h. The solution was centrifuged for 1 h at 100,000 x g, 4 °C. The supernatant contains the SMALPs with $F_oF_1$-mNG-17-mNG-ATP synthase. SMA buffer was used for dilution, if necessary.

### 2.5 Confocal ABEL trap setup with 491 nm laser

Our implementation of a fast FPGA-based ABEL trap was described earlier[6, 35-38, 44, 45] and was slightly modified here. Briefly, a linearly polarized continuous-wave laser emitting at 491 nm (Calypso, 50 mW, Cobolt) was attenuated to 40 µW. Laser beam steering was achieved by a pair of electro-optical beam deflectors, EOBDs (Model 310A, Conoptics). A dichroic beam splitter (H 488 LPXR, AHF, Tübingen, Germany) reflected the laser towards the 60x oil immersion objective (PlanApoN, NA 1.42, Olympus) mounted on an Olympus IX71 inverted microscope. After passing a 200 µm pinhole (Thorlabs), two single photon-counting APDs (SPCM-AQRH-14, Excelitas, Canada) recorded the fluorescence photons in two spectral ranges from 500 to 570 nm (HQ535/70, AHF) and for wavelength $\lambda > 605$ nm (594 LP EdgeBasic, AHF), separated by a dichroic mirror at 588 nm (T 585 LPXR, AHF). Both APDs and the pinhole were mounted on a single 3D-adjustable mechanical stage (OWIS, Germany). Photons were recorded in parallel on two separate computers. One computer contained the FPGA card (PCIe-7852R, National Instruments) used for the feedback of the ABEL trap, and the second computer was connected to an external TCSPC card (SPC-180NX, Becker&Hickl, Germany) in a PCIe box (RAZER Core X) *via* a Thunderbolt 3 cable. The FPGA Labview program from A. Fields and A. E. Cohen[46] was adapted for trapping on the signals of either one or both of the APDs, and a 3D piezo scanner (P-527.3CD with digital controller E-725.3CD, Physik Instrumente, Germany) was implemented in the FPGA software. The microfluidic PDMS chip design we used was published previously[34, 47, 48]. To fabricate the disposable PDMS/glass chips for the ABEL trap, the Sylgard 184 elastomer kit (Dow Corning, Farnell, Germany) was used. Short plasma treatment of both the PDMS chamber and the cover glass ensured irreversible bonding.

## 3. RESULTS

### 3.1 Fluorescent Lifetime Imaging of NTSR1-mNG in living HEK293T cells

To investigate the G protein-coupled receptor NTSR1 in living HEK293T FlpIn cells, we expressed a human NTSR1 mutant tagged with the fluorescent protein mNeonGreen, NTSR1-mNG, at the C-terminus. First we checked the expression of the NTSR1 mutant by confocal laser scanning microscopy under physiological conditions. NTSR1-mNG was excited with a ps-pulsed 488 nm laser at a repetition rate of 40 MHz using a STED microscope from Abberior Instruments. Stably transfected HEK cells were grown and measured in a microfluidic chamber with defined glass bottom for high-resolution imaging. After an 1 h starvation period in PBS+ buffer (i.e., PBS supplemented with $Ca^{2+}$ and $Mg^{2+}$) at 37 °C, NTSR1-mNG was localized specifically in the plasma membrane of the HEK cells (Fig. 2A, left image). Fluorescence intensity from the plasma membrane appeared to be comparable between the different adjacent cells in the field of view. The homogeneous brightness along the membranes of individual cells indicated an average NTSR1-mNG concentration which was too high for single-molecule detection or for identification of individual receptors, respectively.

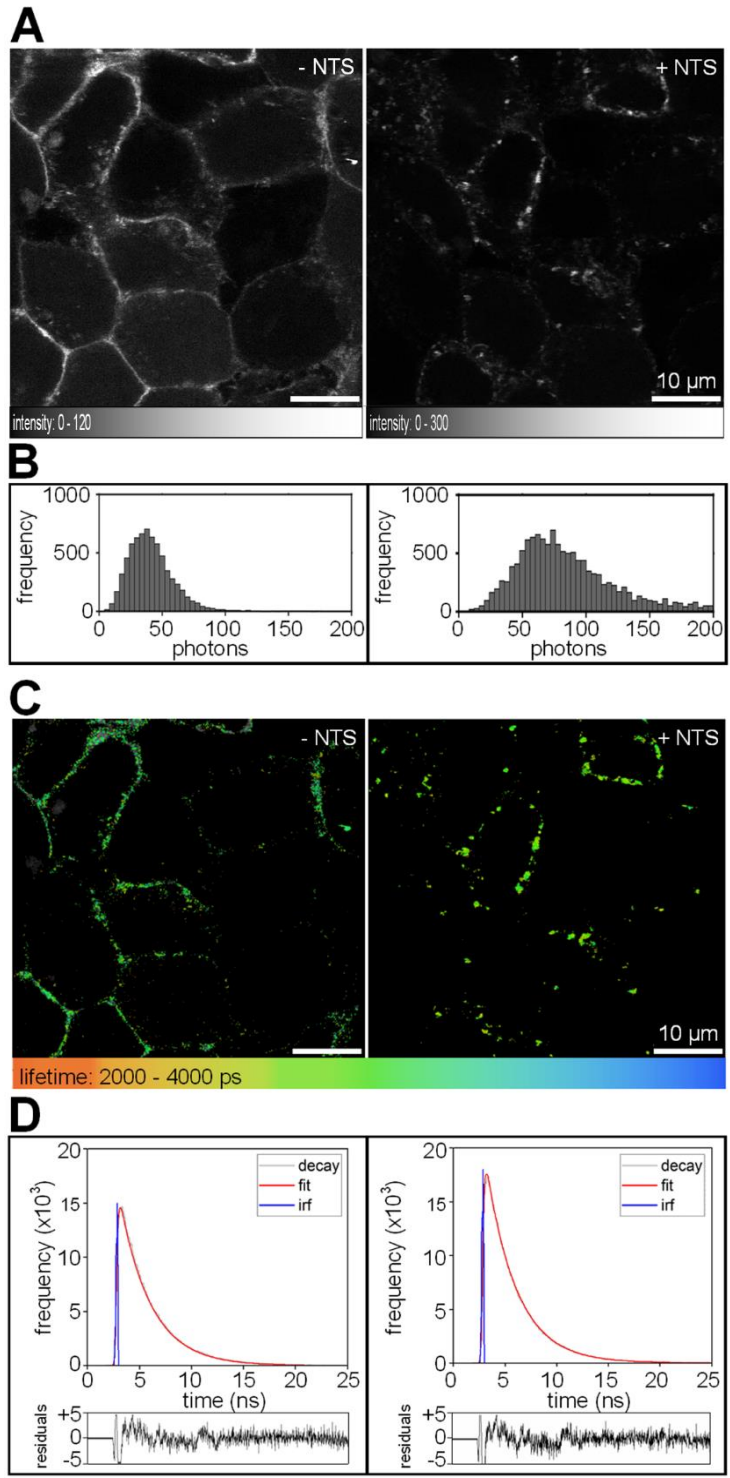

**Figure 2**: confocal FLIM of NTSR1-mNG expressed in HEK293T FlpIn cells at 37 °C before (left panels) and after (right panels) addition of the ligand NTS. (A) normalized fluorescence intensity images with related gray scale maps, (B) pixel intensity histograms, (C) fluorescence lifetime maps, (D) selected pixel-combined fluorescence lifetime decays (gray curves) with instrument response function, irf (blue curves), fitted decays (red curves) and residuals (below).

To visualize the plasma membranes with high contrast, we chose a confocal z-plane position few µm above the cover glass. However, according to 3D z-stacks the plasma membranes of adjacent HEK cells were not arranged orthogonally to the x,y planes of the confocal scans, but mostly tilted by varying angles (data not shown). This resulted in an apparent broadening of the plasma membrane in the confocal image using an 60x water immersion objective in combination with a pinhole set to one Airy unit. Already before addition of the ligand neurotensin, larger spots of NTSR1-mNG were found in the inner parts but near the plasma membranes of some HEK cells.

Next we added the 13 amino acid peptide neurotensin, NTS, at 37 °C and continued to record the same set of HEK cells (Fig. 2A, right image). After addition of its ligand, an aggregation of NTSR1-mNG was observed within a few minutes. The receptor clusters were located in or near the plasma membranes. However, time lapse imaging revealed an internalization of these NTSR1-mNG clusters into the cells after some additional time, i.e., within one hour after addition of NTS (data not shown).

When we compared the two confocal images, we noticed that the maximum pixel brightness was about a factor of 2 different for the conditions before and after addition of neurotensin (note the different intensity scale bars below each Fig. 2A left and right). Accordingly, the NTSR1-mNG clusters were significantly brighter compared to the plasma membranes before stimulation. The pixel intensities were plotted in histograms in Fig. 2B. While a large number of pixels with lower counts were belonging to cytosolic parts of the cells outside of the plasma membrane regions, a larger fraction of pixels with counts > 100 photons was found after addition of NTS and the clustering of NTSR1-mNG.

The identity of mNeonGreen as the bright fluorophore in these confocal images was confirmed by fluorescence lifetime (FLIM) analysis as shown in Fig. 2C. We selected the plasma membranes and the receptor clusters from intracellular background by intensity thresholds and calculated the fluorescence lifetimes in each of the selected pixels using the SPCImage software (see above). The similar green coloring in both false-colored images indicated an apparently unchanged fluorescence lifetime of NTSR1-mNG before and after NTS stimulation. For a more precise fitting of the fluorescence decays, we combined the fluorescence decay curves from all selected pixels into one histogram each (Fig. 2D). For both conditions, the fluorescence lifetimes could be fitted by a monoexponential decay function yielding similar lifetimes $\tau$ = 3.0 ns. Aggregation of NTSR1-mNG by NTS binding did not alter the fluorescence lifetime of the marker protein.

## 3.2 Single-molecule spectroscopy of SMALPs with NTSR1-mNG recorded in an ABEL trap

The confocal FLIM recordings of NTSR1-mNG in living HEK293T FlpIn cells did provide qualitative evidence on the increasing oligomerization state of the receptor after binding its ligand. However, confocal imaging did not provide information to discriminate a monomeric from a possible dimeric state of NTSR1-mNG before NTS addition, nor it provided information about the cluster size of the aggregated NTSR1-mNG after NTS binding.

Therefore, we aimed at trapping the oligomerization state of NTSR1-mNG in the membranes by fast fragmentation of all cellular lipid bilayers into soluble, small lipid nanodiscs with diameters of less than 20 nm. These nanodiscs were expected to accommodate the receptor in different oligomerization states as well as all other membrane proteins. To produce the membrane nanodiscs, we added 2% of a styrene-maleic acid copolymer SMA to the HEK293T cell layer in the presence of benzonase, an efficient nuclease which depolymerized all DNA and RNA. The process of SMALP preparation from living HEK293T cells is shown in Fig. 3A. Immediately after addition of SMA, the cells dissolved into an almost transparent solution. Removing the remaining cellular debris and insoluble large particles by ultracentrifugation allowed to harvest the SMALPs. Fluorescence of the soluble SMALPs was caused by mNeonGreen according to fluorescence lifetime analysis. A monoexponential decay fit yielded a lifetime of $\tau \sim 3$ ns as expected for mNeonGreen. The small size of the SMALPs with NTSR1-mNG, i.e., less than 20 nm on average and with a narrow size distribution, was confirmed by diffusion measurements using fluorescence correlation spectroscopy (data not shown). Diffusion times of SMALPs were about tenfold larger than those of freely diffusing single fluorophores like Alexa Fluor 488 or ATTO 488 in water, i.e., corresponding to a tenfold larger hydrodynamic radius compared to single fluorophores with a hydrodynamic radius of about 1nm. However, FCS and lifetime analysis did not provide a direct information of the oligomerization states of NTSR1-mNG in SMALPs.

To enable a discrimination of monomeric from dimeric or higher oligomeric NTSR1-mNGs in SMALPs by fluorescence intensities we applied single-molecule trapping of these membrane nanodiscs in solution using a fast confocal anti-

Brownian electrokinetic trap, ABEL trap (Fig. 3). The design of the PDMS/glass microfluidic chip limited SMALP diffusion to two dimensions (x and y with respect to the microscope stage) due to a constricted height of about 1 μm in the trapping region (z dimension). Fast moving of the laser focus across the 2.3 x 2.3 μm$^2$ trapping region in an 32-point knight-tour pattern achieved the homogeneous spatio-temporal illumination yielding a constant fluorescence intensity of a trapped SMALP (Fig. 3B). The knight-tour pattern was generated at a 20 kHz repetition rate. Setting the ABEL trap software parameters for the estimated diffusion coefficient to D = 40 μm$^2$/s and for the estimated electromobility to μ = -800 μm/s/V resulted in efficient trapping of the SMALPs. Thus the feedback controls of the ABEL trap yielded increased observation times of individual SMALPs up to 1.4 s, i.e., about a thousand times longer than for freely diffusing SMALPs.

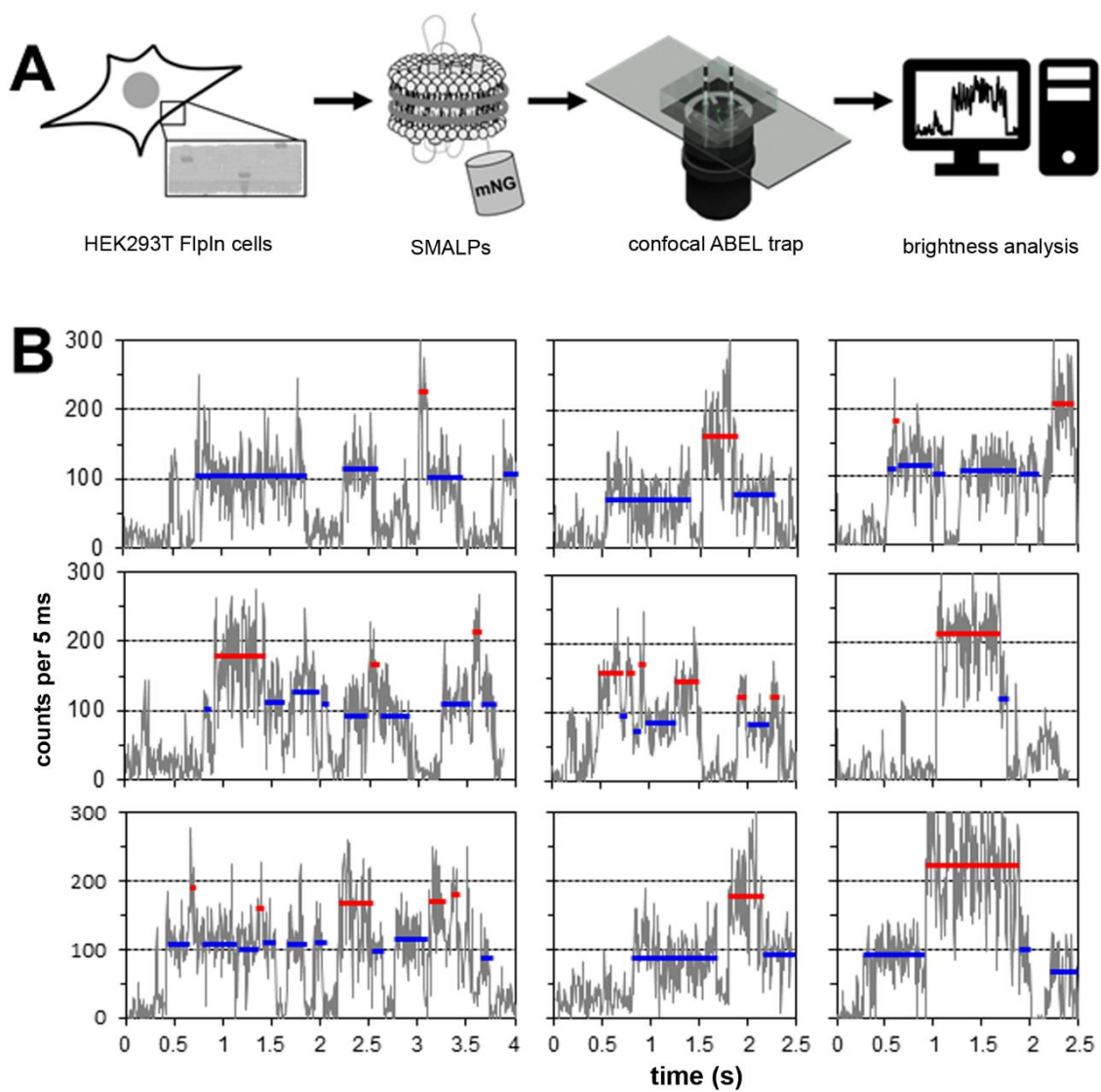

**Figure 3**: (A) Scheme for single-molecule spectroscopy of NTSR1-mNG in SMALPs produced from HEK293T FlpIn cells. (B) Intensity time traces of single SMALPs with activated NTSR1-mNG hold in solution by a confocal ABEL trap. HEK293T FlpIn cells expressing NTSR1-mNG were incubated with the ligand neurotensin before dissolving by the addition of styrene-maleic acid copolymer. Continuous-wave laser excitation at 491 nm, fluorescence detection for 500 nm < λ < 570 nm, photon counts by 5 ms binning, background subtracted. Two distinct brightness levels (blue and red lines) were discerned.

Examples from the brightness measurements of individual trapped SMALPs containing NTSR1-mNG after stimulation with NTS are shown in Fig. 3B. Photons were binned to 5 ms intervals in the intensity time traces. Two distinct brightness levels were found suggesting that the SMALPs contained either one or two NTSR1-mNG.

The blue-marked intensity levels in Fig 3B were assigned to the brightness of a single mNeonGreen fluorophore. The mean brightness of about 100 photons per 5 ms time bin corresponded to a count rate of 20 kHz. This brightness was attributed previously to a single mNeonGreen fluorophore that was fused to the membrane protein $F_oF_1$-ATP synthase and was recorded in our ABEL trap[37]. The red-marked intensity levels in Fig. 3B with about twice the intensity of a single mNeonGreen were interpreted as the brightness of two mNeonGreen fluorophores, or as two NTSR1-mNGs in a single SMALP, respectively. Fluctuations between the two distinct brightness levels during the observation time of one trapped SMALP could be caused by blinking or by photobleaching of one mNeonGreen fluorophore. We noticed that the brightness levels for both a monomer or a dimer of NTSR1-mNG in SMALPs were not identical but varied from SMALP to SMALP, and sometimes changed within the observation time a single SMALP.

Fig. 4 summarizes the brightness levels found in three ABEL trap data sets recorded for 25 min each. Trapped SMALPs were selected by manually marking the start and the end of a photon burst. After background subtraction, mean photon count rates for discrete brightness levels within the selected SMALPs were calculated separately. The duration of each brightness level was noted. Brightness levels with less than 70 counts per 5 ms bin or with more than 400 counts per 5 ms bin were excluded. Photon bursts with short durations of less than 100 ms or with strong intensity fluctuations were excluded.

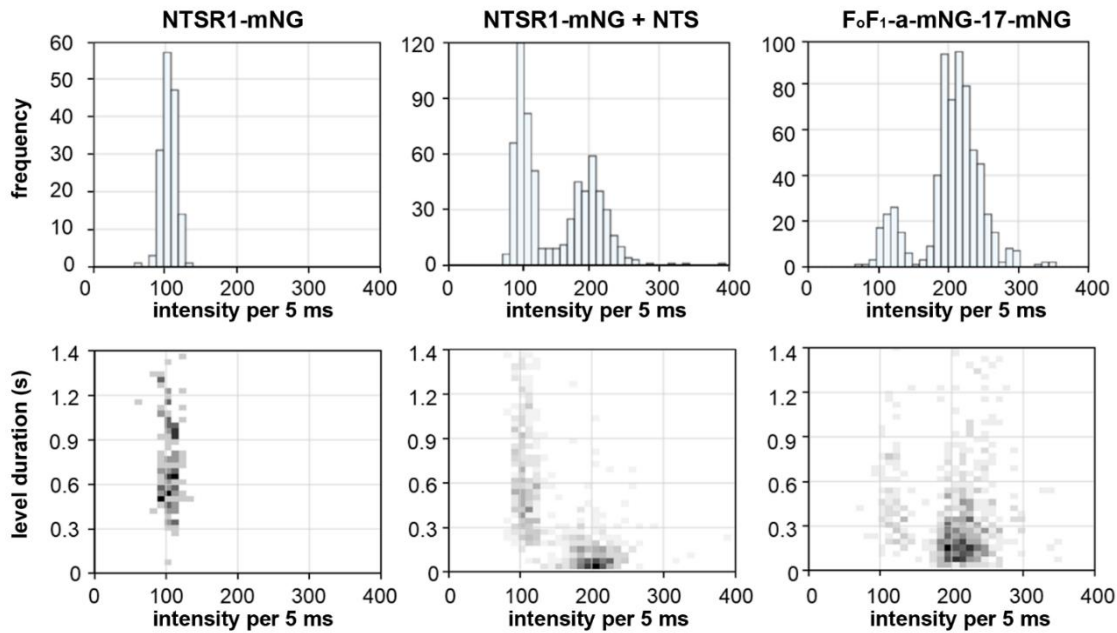

**Figure 4**: Brightness level distributions of single SMALPs with NTSR1-mNG before (one- and two-dimensional histograms on the left) and after stimulation with NTS (one- and two-dimensional histograms in the middle). The brightness level reference of SMALPs with $F_oF_1$-ATP synthase comprising a covalently linked mNeonGreen dimer fused to the *a*-subunit is shown in the one- and two-dimensional histograms on the right. Photon counts were binned to 5 ms intervals.

SMALPs containing NTSR1-mNG before addition of NTS showed a single and narrow brightness distribution, with mean intensity levels of 100 counts per 5 ms bin. Durations of brightness levels ranged from 0.3 to 1.4 s (Fig. 4, left side). In contrast, SMALPs with NTSR1-mNG after stimulation with the ligand NTS showed two distinct populations of brightness levels (Fig. 4, middle). The mean intensity of the additional population was found around 200 counts per 5 ms, i.e., at twice the intensities of SMALPs with NTSR1-mNG before addition of NTS. However, the mean level duration for these bright levels was significantly shorter and most of the brighter levels lasted for less than 200 ms.

To compare the brightness levels found for SMALPs with NTSR1-mNG, we produced SMALPs from *E. coli* expressing the membrane-bound enzyme $F_oF_1$-ATP synthase. The $F_oF_1$-ATP synthase mutant was tagged with a permanent mNeonGreen dimer at the C-terminus of the *a*-subunit ($F_oF_1$-a-mNG-17-mNG). The two mNeonGreen fluorophores were covalently linked by a 17 amino acid linker (see Materials and Methods). The brightness level distributions of SMALPs with $F_oF_1$-a-mNG-17-mNG are shown in Fig. 4 on the right side. Two populations were found with intensities around 120 and 220 counts per 5 ms bin. Most brightness levels were found for the population around 220 counts per 5 ms bin, with level durations ranging from 0.3 to 1.4 s. The population with intensities around 120 counts per 5 ms exhibited a similar distribution of brightness level durations. Because the enzyme $F_oF_1$-ATP synthase is expected to act as a monomer but not as a dimer in membrane of *E. coli*, we assigned to bright population with 220 counts per 5 ms bin to represent a dimeric mNeonGreen fluorophore in a SMALP as the reference. We noticed the slightly higher mean brightness levels of the two $F_oF_1$-a-mNG-17-mNG populations with respect to NTSR1-mNG in SMALPs.

## 4. DISCUSSION

Activation and internalization of the G protein-coupled receptor NTSR1 had been reported after transfection and expression in COS-7 cells lacking an endogenous NTSR1[27, 49]. Time-dependent colocalization of a BODIPY-labeled neurotensin ligand and an immunolabeled NTSR1 inside the cells was revealed using confocal microscopy. NTSR1 and NTS were found in acidic endosomes, i.e., most likely in lysosomes. About 20 min after internalization, the fluorescent ligand had been dissociated from NTSR1 in these endosomes.

We chose HEK293T FlpIn cells for stable expression of human NTSR1 mutants that were fluorescently marked at the C-terminus by genetic fusion with fluorescent proteins. Previously, we measured a human NTSR1 tagged with mRuby3 in HEK293T FlpIn cells[6, 26]. Time-lapse wide-field 3D-recording of NTSR1-mRuby3 confirmed the plasma membrane localization of the receptor in the absence of NTS as well as the NTS-induced internalization in living cells at physiological conditions. However, cellular fluorescence of NTSR1-mRuby3 was too dim for structured illumination microscopy. The marker mRuby3 exhibits a high maximum absorbance coefficient $\varepsilon$=128,000 $M^{-1}\cdot cm^{-1}$ at 558 nm (see www.fpbase.org) compared to the reference fluorescent protein eGFP (maximum $\varepsilon$=56,000 $M^{-1}\cdot cm^{-1}$ at 489 nm; www.fpbase.org), but its reported fluorescence quantum yield is only QY=0.45[25]. In contrast, the fluorescent protein mNeonGreen exhibits a similarly high absorbance coefficient (maximum $\varepsilon$=116,000 $M^{-1}\cdot cm^{-1}$ at 505 nm; www.fpbase.org) combined with a much higher QY=0.8[24]. Therefore, we expected that mNeonGreen as the marker fused to the C-terminus of the NTSR1, i.e., NTSR1-mNG, should provide an optimal signal-to-noise ratio for imaging and for single-molecule detection[3].

Here we compared the distributions of the human NTSR1-mNG in HEK293T FlpIn cells before and after addition of its peptide ligand NTS. Our goal was to identify fluorescence microscopy methods that allow for the discrimination between a monomeric, a dimeric or an oligomeric state of the receptor before and in response to the binding of NTS. First, confocal microscopy was evaluated in combination with recording of fluorescence lifetimes (FLIM). NTSR1-mNG in living HEK293T FlpIn cells was measured in thin microfluidic chambers for high-resolution imaging at physiological conditions, i.e., at 37 °C. A starvation period for up to two hours using a minimal PBS+ buffer instead of the growth medium resulted in the localization of most NTSR1-mNG in the plasma membranes of the HEK cells. The membranes appeared homogeneously fluorescent indicating that (a) the NTSR1-mNG concentration is significantly higher than one receptor per $\mu m^2$, and/or (b) the mobility of NTSR1-mNG is high to induce motion blur during the confocal laser scanning.

Excitation was performed with a picosecond-pulsed laser of a fast laser scanning microscope and fluorescence was detected with single photon counting APDs and TCSPC electronics. This allowed for fluorescence lifetime calculations to prove the fluorophore's identity. FLIM revealed a lifetime $\tau$=3.0 ns at 37 °C, i.e., the same lifetime for NTSR1-mNG in HEK cells at 21 °C we had reported previously using our custom-built confocal microscope with slow piezo scanning and large pixel sizes[26].

Addition of NTS to the HEK cells caused the formation of bright fluorescence spots at the membrane within 10 minutes, or NTSR1-mNG clustering, respectively, at physiological temperatures of 37 °C. Comparing the maximum pixel brightness before and after clustering revealed an increase by a factor 2. The fluorescence lifetime remained the same with $\tau$ = 3.0 ns as found previously in our measurements at 21 °C.

For NTSR1-mNG imaging in living HEK cells at 37 °C we used a modular STED microscope (Expertline, Abberior Instruments) comprising eight pulsed lasers and six APDs. These APDs enable polarization-resolved FLIM recordings in three spectral ranges in parallel. Static and time-resolved fluorescence anisotropy imaging could reveal a multimeric state of NTSR1 through homoFRET, i.e., Förster resonance energy transfer between identical and nearby fluorophores, and homoFRET is detected by a smaller fluorescence anisotropy. Using our confocal piezo scanning microscope for anisotropy imaging at 21 °C we previously noticed small changes of fluorescence anisotropy distributions of NTSR1-mNG before and after addition of NTS[26]. For NTSR1-mNG anisotropy imaging in living HEK cells at 37 °C anisotropy calibrations and data analyses are still in progress.

The NTS-induced formation of aggregated NTSR1-mNGs as densely packed membrane proteins suggested the possibility that receptor oligomers could have been formed. In order to calculate the composition of the putative oligomers and to discriminate a monomeric *versus* a dimeric oligomerization state of NTSR1-mNG before ligand activation, we dissolved all cellular membranes into lipid nanodiscs by adding a SMA copolymer (XIRAN SL25010 P20, Polyscope). The process of SMALP preparation from living HEK293T cells is shown in Fig. 1. The resulting soluble SMALPs with a diameter of approximately 15 nm were separated from cellular debris by ultracentrifugation. After a 100- to 150-fold dilution to adjust fluorescent nanoparticle concentrations between 20 pM and 100 pM, the fluorescence brightness of individual SMALPs were recorded on a confocal ABEL trap one at a time. ABEL trapping of single biomolecules had been used for counting bound ATP molecules by stepwise photobleaching[50], conformational changes of a GPCR[51] and photophysical properties of photosynthetic proteins by lifetime[52, 53], reversible binding and aggregation by diffusion and electromobility analysis[46, 54, 55] and precise intramolecular distance measurements by smFRET[38, 48, 56].

In the absence of NTS, a well-defined brightness level was found for SMALPs comprising NTSR1-mNG, other membrane proteins and lipids. Given an excitation power of 40 mW measured in the laser beam path before the microscope objective, a mean photon count rate of 20 kHz was attributed to the monomer of NTSR1-mNG or a single mNeonGreen marker fluorophore, respectively. A similar mean brightness was found for our reference membrane protein $F_oF_1$-ATP synthase tagged with one mNG on the *a*-subunit[37, 40] and reconstituted as a single enzyme into liposomes. We note that thresholds were applied to select the photon bursts manually, i.e., a minimum trapping duration of 100 ms, a minimum mean brightness of 14 kHz and a maximum mean brightness of 80 kHz. Photon bursts with fast fluctuating intensity level were excluded. Trapping times of the NTSR1-mNG SMALPs reached up to 1.4 s, i.e., a thousandfold increase in observation times compared to freely diffusing SMALPs (Fig. 4).

After adding NTS to HEK cells and inducing receptor internalization for 1 h, the brightness distribution of individual SMALPs revealed two separate fluorescence intensity level with mean photon count rates of 20 kHz and 40 kHz (Fig. 4). We note that the same intensity thresholds were applied as for NTSR1-mNG SMALPs in the absence of NTS. However, durations of less than 100 ms were taken into account for assigning a 40 kHz brightness level. Within the photon burst of an individual trapped SMALP, the intensity level sometimes switched reversibly (Fig. 3). While the majority of the 40 kHz level dropped to the lower brightness within 300 ms, a few of these mean bright intensity level lasted up to 1 s, clearly indicating a stable dimeric NTSR1-mNG in one SMALP.

In addition to NTSR1-mNG SMALPs, we conducted two control experiments. First, we recorded SMALPs containing the mitochondrial $F_oF_1$-ATP synthase tagged on the γ-subunit with mNeonGreen[39] (ABEL trap data not shown). Second, the membrane enzyme $F_oF_1$-ATP synthase tagged with two consecutive, covalently linked mNeonGreen fluorophores on the *a*-subunit was expressed in *E. coli* and named $F_oF_1$-a-mNG-17-mNG. SMALPs were produced from the bacterial membranes to provide a reference for the brightness of a pure mNeonGreen dimer. Although all $F_oF_1$-ATP synthase mutants in SMALPs should exhibit the brightness of a mNeonGreen dimer (corresponding to the mean brightness of 40 to 45 kHz in Fig. 4), a smaller fraction related to the brightness of a single mNeonGreen fluorophore was found as well (mean brightness 20 to 25 kHz, Fig. 4). Similarly to the dimer of NTSR1-mNG in SMALPs, the durations of most of the 40 to 45 kHz brightness level of $F_oF_1$-a-mNG-17-mNG were shorter than 500 ms, but a few long-lasting bright level were found as well.

As the brightness distribution of single $F_oF_1$-a-mNG-17-mNG in SMALPs represented the dimer of mNeonGreen (with a small fraction of one nonfluorescent mNeonGreen in the dimer), the brightness distributions of NTSR1-mNG in SMALPs before and after activation by NTS could be interpreted. Before addition of NTS, the NTSR1-mNG appeared to be monomeric in the plasma membranes of HEK293T FlpIn cells. After addition of NTS and followed by clustering and internalization, almost half of the NTSR1-mNG receptors appeared to be trapped as a dimer within a single SMALP.

We assumed that the presence of the dim 20 kHz brightness level could either be interpreted as evidence for monomeric single NTSR1-mNG trapped in the lipid nanodisc, or as an indication for a nonfluorescent, nonmatured mNeonGreen marker which could not be detected by our microscopy approach. Comparing the different ratios of monomer to dimer populations of NTSR1-mNG after NTS stimulation and $F_oF_1$-a-mNG-17-mNG indicated that not all NTSR1-mNG are part of a dimer.

The SMALP approach to trap the oligomerization state of NTSR1 in living HEK293T FlpIn cells before and after activation could be considered as an alternative to microscopic approached in living cells, and single-nanoparticle analysis using the confocal ABEL trap could reveal the dynamics of monomer and oligomer fractions of the receptor. However, the relative differences in the SMALP brightness histograms have to be revised by controlling the expression numbers of NTSR1 in the cell membranes, and by excluding possible dependencies of the brightness distributions on the lengths and charge of the SMA or similar copolymers, the incubation conditions of the cells and the subsequent ultracentrifugation steps. Then, anisotropy measurements of individual trapped NTSR1-mNG SMALPs could reveal the relative orientations of the markers in dimeric NTSR1-mNG and could distinguish between static or transient relative orientations of the NTSR1-mNG dimers or even receptor oligomers.

**Acknowledgements**

The authors thank all members of our research groups who participated in various aspects of this work, from genetics and biochemistry support to cell culture. Financial support for the Abberior STED superresolution microscope by the State of Thuringia (TMWWDG, "Kapitel 1820, Titel 891 03") and by the Deutsche Forschungsgemeinschaft DFG ("INST 1757/25-1 FUGG", to M.B.), for A.W. through LOM funds 2017 by the Jena University Hospital (to M.B.) and in part by the Deutsche Forschungsgemeinschaft DFG in the Collaborative Research Center/Transregio 166 "ReceptorLight" (project A1 to M.B.) is gratefully acknowledged. The ABEL trap was realized by additional DFG funds (grants BO1891/10-2, BO1891/15-1, BO1891/16-1, BO1891/18-2 to M.B.) and was supported by an ACP Explore project (M.B. together with J. Limpert) within the ProExcellence initiative ACP2020 from the State of Thuringia.